# Quantization Audio Watermarking with Optimal Scaling on Wavelet Coefficients

S.-T. Chen, H.-N. Huang, and S.-Y. Tu

*Abstract*—In recent years, discrete wavelet transform (DWT) provides an useful platform for digital information hiding and copyright protection. Many DWT-based algorithms for this aim are proposed. The performance of these algorithms is in term of signal-to-noise ratio (SNR) and bit-error-rate (BER) which are used to measure the quality and the robustness of an embedded audio. However, there is a tradeoff relationship between the embedded-audio quality and robustness. The tradeoff relationship is a signal processing problem in the wavelet domain. To solve this problem, this study presents an optimization-based scaling scheme using optimal multi-coefficients quantization in the wavelet domain. Firstly, the multi-coefficients quantization technique is rewritten as an equation with arbitrary scaling on DWT coefficients and set SNR to be a performance index. Then, a functional connecting the equation and the performance index is derived. Secondly, Lagrange Principle is used to obtain the optimal solution. Thirdly, the scaling factors of the DWT coefficients are also optimized. Moreover, the invariant feature of these optimized scaling factors is used to resist the amplitude scaling. Experimental results show that the embedded audio has high SNR and strong robustness against many attacks.

## I. INTRODUCTION

Generally, an audio watermarking scheme has the following requirements [1], [2]: (1) The watermarks should be imperceptible in the embedded audio. To ensure this requirement, the embedding technique should offer at least 20dB signal to noise ratio (SNR); (2) The

S.-T. Chen is with the Department of Mathematics, Tunghai University, Taichung 40704, TAIWAN (corresponding author; e-mail: shough34@yahoo.com.tw).
H.-N. Huang is with the Department of Mathematics, Tunghai University, Taichung 40704, TAIWAN (e-mail: nhuang@thu.edu.tw).
S.-Y. Tu is with the Mathematics Department, University of Michigan-Flint, Flint MI 48502, USA (e-mail: sytu@umflint.edu).

embedded watermarks should be able to prevent common attacks, such as re-sampling, MP3 compression, filtering, amplitude scaling and time scaling.

Most methods for audio watermarking can be grouped into two categories, time-domain technique [3-12] and frequency-domain technique [2], [13-18]. In the time-domain technique, Lie *et al.* [7] adopted the amplitude modification to improve robustness in the time domain. However, it has an extremely low capacity and SNR. This is because they use three segments (length = 1020 points) to present one bit. In the frequency-domain technique, Huang *et al.* [13] embeds the watermark into discrete cosine transform (DCT) coefficients, and hides the Bar code in the time domain as synchronization codes. Because the time domain has low embedding strength, the synchronization codes are not robust enough. However, if the synchronization codes are embedded in DCT, then the computation cost increases. Generally, wavelet-based watermarking has good performance. To overcome the drawback in the method proposed by Huang *et al.* [13] in the wavelet domain, Wu *et al.* [2] used quantization index modulation method to embed synchronization codes and watermarks into low-frequency coefficients in DWT. Their method achieves better robustness against common signal processing and noise corruption. However, it is very vulnerable to amplitude and time scaling because of the single coefficient quantization. Xiang *et al.* [16] implemented the method proposed by Lie *et al.* [7] in the wavelet domain. Their method has slightly improved results.

In the field of audio watermarking, the embedded-audio quality and robustness are typically measured by signal-to-noise ratio (SNR) and bit error ratio (BER). As far as these two measurement tools are concerned, there is a tradeoff relationship between them. Chen *et al.* [18] proposed an optimization-based scheme to obtain the best embedded-audio quality under the fixed scaling on DWT coefficients which showed that the hidden data are robust against some common attacks. However, the embedded-audio quality decreases when the scaling factors have a variation and the embedded watermarks are not sufficiently robust to amplitude scaling. In this paper, the optimization-base scheme is extended to include the scaling of DWT coefficients to find the best embedded-audio quality. The novel embedding technique is firstly rewritten as an equation with arbitrary scaling on DWT coefficients and the SNR is set to be a performance index which is a function of DWT coefficients. By integrating these two terms into a new one which is a function of function, a novel wavelet-based functional connecting the equation and the performance index is derived. Secondly, Lagrange Principle is used to derive the optimal conditions. Thirdly, the optimal scaling factors of the DWT coefficients are obtained under the sense of minimum-length solution. Moreover, the invariant feature of these optimal scaling factors is used to resist the amplitude scaling. Based on the above theoretical results, this study presents an optimization-based scaling scheme using optimal multi-coefficients quantization in the wavelet domain.

The rest of this paper is organized as follows. Section II reviews DWT and introduces the optimization-based scaling scheme for embedding and extraction. Section III derives a wavelet-based functional that connects the multi-coefficients quantization equation and the performance index. The Lagrange Principle is used to obtain the optimal DWT coefficients. Moreover, the scaling factors of DWT coefficients are also optimized in this section. Section IV does some experiments to test the performance. Conclusions are finally drawn in Section V.

## Ⅱ. THE PROPOSED EMBEDDING AND EXTRACTION TECHNIQUES

Unlike the traditional single–coefficient quantization in [2], the novel amplitude quantization technique with optimal scaling is proposed in this section. Before the proposed technique, DWT is reviewed as follows.

*A. Discrete wavelet transforms*

The wavelet transform is obtained by a single prototype function $\psi(x)$ which is regulated with scaling parameter and shift parameter [19, 22]. The discrete normalized scaling and wavelet basis function are defined as

$$\varphi_{i,n}(t) = 2^{i/2} h_i(2^i t - n) \tag{1}$$

$$\psi_{i,n}(t) = 2^{i/2} g_i(2^i t - n) \tag{2}$$

where $i$ and $n$ are the dilation and translation parameters; $h_i$ and $g_i$ are lowpass and highpass filters. Orthogonal wavelet basis functions not only provide simple calculation in coefficients expansion but also span $L^2(\mathbb{R})$ in signal processing. As a result, audio signal $S(t) \in L^2(\mathbb{R})$ can be expressed as a series expansion of orthogonal scaling functions and wavelets. More specifically,

$$S(t) = \sum_{\ell} c_{j_0}(\ell)\varphi_{j_0,k}(t) + \sum_{k}\sum_{j=j_0}^{\infty} d_j(k)\psi_{j,k}(t) \qquad (3)$$

where $c_j(\ell) = \int_{\mathbb{R}} S(t)\varphi_{j,\ell}(t)dt$ and $d_j(k) = \int_{\mathbb{R}} S(t)\psi_{j,k}(t)dt$ be the low-pass and high-pass coefficients, respectively; $j_0$ is an integer to define an interval on which $S(t)$ is piecewise constant. This work uses orthogonal basis to implement DWT through filter bank. Fig. 1 shows an example that the digital audio signal $S(n)$ is decomposed into four sub-bands by applying three-level DWT. In this paper, $S(n)$ is transformed into wavelet domain by seven-level decomposition. For the consideration of robustness, we embed the synchronization codes and watermarks into the lowest-frequency sub-band coefficients.

*B. Embedding technique*

In the proposed embedding technique, we first split the original audio into proper segments. Then, DWT is performed on each segment. Since the watermarked audio may suffer the attack of shifting or cropping, it is necessary to embed the synchronization codes together. These synchronization codes are used to locate the positions where the watermark is embedded. The structure is shown in Fig. 2. Before embedding, the synchronization codes and watermarks are arranged into a binary pseudo-random noise (PN) sequence, $B = \{\beta_i\}$, e.g., $B = \{1,0,1,1,\cdots\}$. Then, the binary sequence is embedded into lowest-frequency sub-band of each segment by the following rules:

- If the bit "$1 \in B$" is embedded into $N$ consecutive coefficients $\{|c_1|,|c_2|,\cdots,|c_N|\}$, their group-amplitude $\sum_{j=1}^{N}|c_j|$ is quantized to

$$\gamma_1 = \left\lfloor \frac{\sum_{j=1}^{N}|c_j|}{Q} \right\rfloor Q + \frac{3}{4}Q \qquad (4)$$

- If the bit "$0 \in B$" is embedded into $N$ consecutive coefficients $\{|c_1|,|c_2|,\cdots,|c_N|\}$, their group-amplitude $\sum_{j=1}^{N}|c_j|$ is quantized to

$$\gamma_0 = \left\lfloor \frac{\sum_{j=1}^{N}|c_j|}{Q} \right\rfloor Q + \frac{1}{4}Q \qquad (5)$$

where $\{c_j\}$ are the lowest-frequency coefficients in DWT; $\lfloor \ \rfloor$ indicates the floor function, and $Q$ is the quantization parameter which is adopted as the secret key $K$. Fig. 3 shows the embedding process.

*C. Extraction technique*

The extract technique can extract the watermark without original audio. Similar to embedding, we can split the test audio into segments and then perform DWT on each segment. Let $\bar{C}_N^* = \{|\bar{c}_1^*|,|\bar{c}_2^*|,\cdots,|\bar{c}_N^*|\}$ be the $N$ consecutive coefficients of lowest-frequency sub-band, the binary sequence is extracted from $\bar{C}_N^*$ and optimal scaling factors $a_j$ (which are addressed in later section III-B):

- If

$$\sum_{j=1}^{N} a_j \left|\overline{c}_j^*\right| - \left\lfloor \left. \sum_{j=1}^{N} a_j \left|\overline{c}_j^*\right| \middle/ Q \right. \right\rfloor Q \geq \frac{Q}{2}, \tag{6}$$

the extracted value is 1.

- If

$$\sum_{j=1}^{N} a_j \left|\overline{c}_j^*\right| - \left\lfloor \left. \sum_{j=1}^{N} a_j \left|\overline{c}_j^*\right| \middle/ Q \right. \right\rfloor Q < \frac{Q}{2}, \tag{7}$$

the extracted value is 0.

The detail extraction process is shown in Fig. 4.

### III. THE PROPOSED OPTIMIZATION-BASED EMBEDDING

In this section, some matrix operations are first reviewed and then the proposed optimization-based embedding with optimal scaling on DWT coefficients is presented.

#### A. *Matrix operations and Lagrange Principle*

Some optimization operations from [20], [21] and Lagrange Principle are summarized as follows for calculating the extreme of a matrix function.

**Theorem 1.** *If $A$ is a $n \times n$ matrix, and $\overline{C}$ and $C$ are $n \times 1$ column vectors, then the followings hold:*

$$\frac{\partial A\overline{C}}{\partial \overline{C}} = A, \tag{8}$$

$$\frac{\partial (\overline{C} - C)^T (\overline{C} - C)}{\partial \overline{C}} = 2(\overline{C} - C). \tag{9}$$

The gradient of a matrix function $f(C)$ is defined by

**Definition 1.** Suppose that $C = [c_1, c_2, \cdots, c_n]^T$ is a $n \times 1$ matrix and $f(C)$ is a matrix function. Then the gradient of $f(C)$ is

$$\nabla f(C) = \frac{\partial f}{\partial C} = [\frac{\partial f}{\partial c_1}, \frac{\partial f}{\partial c_2}, \cdots, \frac{\partial f}{\partial c_n}]^T \tag{10}$$

Now we consider the problem of extremizing the matrix function $f(C)$ subjected to a algebraic constraint $g(C) = 0$, i.e.,

$$\textit{minimize} \quad f(C) \tag{11a}$$

$$\textit{subjected to} \quad g(C) = 0 \tag{11b}$$

In order to solve (11), Lagrange Principle given in the following theorem will be applied.

**Theorem 2. (Lagrange Principle)** *Suppose that $g$ is a continuously differentiable function of $C$ on a subset of the domain of a function $f$. If $C_0$ minimizes (or maximizes) $f(C)$ subjected to the constraint $g(C) = 0$, then $\nabla f(C_0)$ and $\nabla g(C_0)$ are parallel. That is, if $\nabla g(C_0) \neq 0$, then there exists a scalar $\lambda$ such that*

$$\nabla f(C_0) = \lambda \nabla g(C_0). \tag{12}$$

Based on Theorem 2, if we define an augmented function as follows

$$H(C, \lambda) = f(C) + \lambda^T g(C), \tag{13}$$

then to find the optimal solution of the constraint problem (11) becomes to compute the extreme of the unconstraint function $H(C, \lambda)$. The necessary conditions for existence of the extreme of $H$ are

$$\frac{\partial H}{\partial \lambda} = 0, \quad \frac{\partial H}{\partial C} = 0. \tag{14}$$

Since there is a tradeoff relationship between the audio quality (SNR) and the robustness (BER), we introduce a scalar parameter $\lambda$ to connect the watermarking cost function and amplitude quantization equation. Finally, Lagrange Principle in Theorem 2 is applied to derive the optimal conditions. And the associated minimum-length solution is computed to obtain the optimal scaling on DWT coefficients.

*B. Optimization problem for the best embedded-audio quality*

Since $\sum_{j=1}^{N}|c_j|$ is quantized to embed the binary bit "$1 \in B$" or "$0 \in B$" according to the procedure defined in Section II-A, we need to determine $N$ unknown values of lowest-frequency DWT coefficients, $\overline{c}_1, \overline{c}_2, \ldots, \overline{c}_N$, together with positive scaling factors, $a_1, a_2, \ldots, a_N$, such that $\sum_{j=1}^{N} a_j |\overline{c}_j| = \gamma_0$ when embedding the bit "$0 \in B$" or $\sum_{j=1}^{N} a_j |\overline{c}_j| = \gamma_1$ when embedding the bit "$1 \in B$". An optimization-based method is proposed to obtain these DWT coefficients as following. Suppose the $N$ unknown absolute values of the lowest-frequency coefficients are put into a vector form

$$\overline{C}_N = \left[|\overline{c}_1|, |\overline{c}_2|, \cdots, |\overline{c}_N|\right]^T \tag{15}$$

with respect to the original DWT coefficient vector $C_N = \left[|c_1|, |c_2|, \cdots, |c_N|\right]^T$, then the embedding technique can be rewritten as

$$A\overline{C}_N = \gamma_1, \text{ if "}1 \in B\text{" is embedded,} \tag{16}$$

or

$$A\overline{C}_N = \gamma_0, \text{ if "}0 \in B\text{" is embedded,} \tag{17}$$

where $A = [a_1 \ a_2 \ \cdots \ a_N]$ is the corresponding scaling matrix whose entries are positive and can be arbitrarily assigned by an encoder. To avoid the situation that the value of some entries become arbitrarily large, without loss of generality we may set the summation of all the scaling factors equal to a constant $M$. For example, $A = [0.9 \ 1.2 \ 1.2 \ 0.7]$ is a suitable selection when $N = 4$ and $M = 4$.

Next step, we consider the optimization problem for watermarking is to select the vector $\overline{C}_N$ such that the SNR is maximized under the constraint (16) or (17). The SNR is calculated as follows

$$SNR = -10 \log_{10} \left( \frac{\|\overline{S}(n) - S(n)\|_2^2}{\|S(n)\|_2^2} \right) = -10 \log_{10} \left( \frac{\|\overline{C}_N - C_N\|_2^2}{\|C_N\|_2^2} \right) \tag{18}$$

This is because we implement the DWT with orthogonal wavelet bases. To understand the effect of adjusting scaling factors on the SNR, we consider the case with $N = 4$, $M = 4$ and the scaling matrix to be $A = [a_1 \ a_2 \ 1 \ 1]$, i.e., $a_1 + a_2 = 2$. Fig. 5 shows the relationship between the SNR and the scaling factor $a_2$ for audio love song, symphony, dance, and folklore, respectively. Detail information about each type of song is described in Section V. As the scaling factor $a_2$ is decreasing from one, the SNR is decreasing as well as the SNR reaches its maximal value when the scaling factor $a_2$ is one (i.e., both scaling factors are equal to one). Therefore it is interesting to find the scaling factors such that the SNR is maximized.

For optimization purpose, we can define the performance index from (18) by

$$\frac{\left\| \overline{C}_N - C_N \right\|^2}{\left\| C_N \right\|^2} \tag{19}$$

or equivalently,

$$\frac{(\overline{C}_N - C_N)^T (\overline{C}_N - C_N)}{C_N{}^T C_N}. \tag{20}$$

This is due to the fact that logarithmic function is a one-to-one function. Since $C_N{}^T C_N$ is a constant, we can express the performance index in (20) in the form as

$$(\overline{C}_N - C_N)^T (\overline{C}_N - C_N) \tag{21}$$

As for the case to embed the bit "$1 \in B$", the optimization-based quantization problem can be described as following:

$$\text{minimize} \quad (\overline{C}_N - C_N)^T (\overline{C}_N - C_N) \tag{22a}$$

$$\text{subjected to} \quad \text{(a)} \ A\overline{C}_N = \gamma_1, \tag{22b}$$

$$\text{(b)} \ \sum_{i=1}^{N} a_i = M. \tag{22c}$$

This optimization problem can be solved in two steps: the first step is to find the optimal solution to (22a) and (22b), and the second one is to adjust this optimal solution such that (22c) is simultaneously satisfied.

As shown in the proposed embedding process, see Fig. 1, to embed the binary bit "1", we need to solve (22). By Theorem 2, we set the Lagrange multiplier $\lambda$ to combine (22a) and (22b) into a scalar function with matrix variable.

$$H(\overline{C}_N, \lambda) = (\overline{C}_N - C_N)^T (\overline{C}_N - C_N) + \lambda^T [A\overline{C}_N - \gamma_1] \tag{23}$$

which has no constraint. The necessary conditions for existence of the minimum of

$H(\bar{C}_N, \lambda)$ are

$$\frac{\partial H}{\partial \bar{C}_N} = 2(\bar{C}_N - C_N) + A^T \lambda = 0 \tag{24a}$$

$$\frac{\partial H}{\partial \lambda} = A\bar{C}_N - \gamma_1 = 0 \tag{24b}$$

Multiplying (24a) by $A$, we observe that

$$2(A\bar{C}_N - AC_N) + AA^T \lambda = 0 \tag{25}$$

Since $A\bar{C}_N = \gamma_1$, equation (25) is rewritten as

$$(\gamma_1 - AC_N) + \frac{1}{2} AA^T \lambda = 0 \tag{26}$$

Hence the optimal solution for the parameter $\lambda$ is

$$\lambda^* = 2(AA^T)^{-1}(AC_N - \gamma_1) \tag{27}$$

Moreover, by substituting (27) into (24a), the optimal solution for modified coefficients is

$$\begin{aligned}\bar{C}_N^* &= C_N - \frac{1}{2} A^T \lambda^* \\ &= C_N - A^T (AA^T)^{-1}(AC_N - \gamma_1)\end{aligned} \tag{28}$$

where the superscript $*$ denotes the optimal result with respect to the corresponding variable.

Based on (28), the binary bit "$1 \in B$" can be embedded by the optimal modified coefficients $\bar{C}_N^*$. In other words, the binary bit "$0 \in B$" is embedded by using $\gamma_0$ instead of $\gamma_1$.

$$\bar{C}_N^* = C_N - A^T (AA^T)^{-1}(AC_N - \gamma_0) \tag{29}$$

Fig. 6 gives the detail embedding process to incorporate the optimal modified coefficients.

By (28) and (29), the encoder can arbitrarily design the scaling matrix $A$ to obtain the corresponding optimal modified coefficients $\bar{C}_N^*$ and then finish the embedding process.

However, there may be an optimal matrix $A$ to maximize SNR. Accordingly, the optimization problem for finding optimal matrix $A$ is described as follows.

$$\text{Minimize} \quad \left(\bar{C}_N^* - C_N\right)^T \left(\bar{C}_N^* - C_N\right) \tag{30a}$$

$$\text{subjected to} \quad \sum_{i=1}^{N} a_i = M \tag{30b}$$

Since the optimal modified coefficients $\bar{C}_N^* = C_N - A^T(AA^T)^{-1}(AC_N - \gamma_1)$ is obtained, the performance index $\left(\bar{C}_N^* - C_N\right)^T \left(\bar{C}_N^* - C_N\right)$ can be rewritten as a function of the matrix $A$:

$$\begin{aligned} f(A) &= \left(\bar{C}_N^* - C_N\right)^T \left(\bar{C}_N^* - C_N\right) \\ &= \left[C_N - A^T(AA^T)^{-1}(AC_N - \gamma_1) - C_N\right]^T \left[C_N - A^T(AA^T)^{-1}(AC_N - \gamma_1) - C_N\right] \\ &= \left[A^T(AA^T)^{-1}(AC_N - \gamma_1)\right]^T \left[A^T(AA^T)^{-1}(AC_N - \gamma_1)\right] \\ &= (AC_N - \gamma_1)^T (AA^T)^{-1} AA^T (AA^T)^{-1}(AC_N - \gamma_1) \\ &= (AC_N - \gamma_1)^T (AA^T)^{-1}(AC_N - \gamma_1) \\ &= \frac{1}{M}(AC_N - \gamma_1)^T (AC_N - \gamma_1) \end{aligned}$$

By writing $f(a_1, a_2, \ldots, a_N) = f(A)$, one obtains

$$f(a_1, a_2, \ldots, a_N) = \frac{1}{M}\left\{[a_1 \ a_2 \ \ldots \ a_N]\begin{bmatrix}|c_1|\\|c_2|\\ \cdot \\ \cdot \\ \cdot \\ |c_N|\end{bmatrix} - \gamma_1\right\}^T \left\{[a_1 \ a_2 \ \ldots \ a_N]\begin{bmatrix}|c_1|\\|c_2|\\ \cdot \\ \cdot \\ \cdot \\ |c_N|\end{bmatrix} - \gamma_1\right\}$$

$$= \frac{1}{M}\left(a_1|c_1| + a_2|c_2| + \ldots + a_N|c_N| - \gamma_1\right)^T \left(a_1|c_1| + a_2|c_2| + \ldots + a_N|c_N| - \gamma_1\right)$$

$$= \frac{1}{M}\left(a_1|c_1| + a_2|c_2| + \ldots + a_N|c_N| - \gamma_1\right)^2$$

Hence, the optimization problem becomes

$$\text{minimize} \quad \frac{1}{M}\left\{a_1|c_1| + a_2|c_2| + \ldots + a_N|c_N| - \gamma_1\right\}^2 \tag{31a}$$

subjected to $\sum_{i=1}^{N} a_i = M$ (31b)

Again, let

$$H = \frac{1}{M}\left\{\sum_{i=1}^{N} a_i |c_i| - \gamma_1\right\}^2 + \lambda(\sum_{i=1}^{N} a_i - M)$$

The necessary condition is

$$\frac{\partial H}{\partial a_i} = \frac{2}{M}|c_i|\left\{\sum_{i=1}^{N} a_i |c_i| - \gamma_1\right\} + \lambda = 0,$$

and then the optimal solutions are given by

$$|c_1| = |c_2| = \ldots = |c_N|$$

or

$$\sum_{i=1}^{N} a_i |c_i| - \gamma_1 = 0, \quad \lambda = 0.$$

Since the first optimal solution requires that all the absolute values of DWT coefficients $c_1, c_2, \ldots, c_N$ are equal which is not always true. Thus the optimal solution can be obtained by solving the following linear system.

$$\begin{cases} a_1|c_1| + a_2|c_2| + \ldots + a_N|c_N| - \gamma_1 = 0 \\ \sum_{i=1}^{N} a_i = M \end{cases} \quad (32)$$

And since the associated Hessian matrix

$$\left[\frac{\partial^2 H}{\partial a_i \partial a_j}\right]_{i,j=1}^{N} = \left[\frac{2}{M}|c_i||c_j|\right]_{i,j=1}^{N}$$

is positive-definite, the scaling factors $a_1, a_2, \ldots, a_N$ obtained from (32) are also sufficient to achieve the minimum solution of (31). To solve (31), one rewrite $a_N = M - \sum_{i=1}^{N-1} a_i$ and then (31) becomes

$$a_1|c_1| + a_2|c_2| + \ldots + \left(M - \sum_{i=1}^{N-1} a_i\right)|c_N| - \gamma_1 = 0$$

i.e.,

$$(|c_1| - |c_N|)a_1 + (|c_2| - |c_N|)a_2 + \ldots + (|c_{N-1}| - |c_N|)a_{N-1} = \gamma_1 - M|c_N|. \quad (33)$$

Since there are $N-1$ unknowns and only one equation, there are infinitely many solutions and it is frequently desirable to compute the solution of minimum Euclidean length solution. Let

$$P = \begin{bmatrix} |c_1|-|c_N| & |c_2|-|c_N| & \cdots & |c_{N-1}|-|c_N| \end{bmatrix},$$
$$x = \begin{bmatrix} a_1 & a_2 & \cdots & a_{N-1} \end{bmatrix}^T, \qquad b = \gamma_1 - M|c_N|,$$

then the matrix form of (33) is $Px = b$. The minimum-length solution is given by $x_+ = P^T v$ where $v$ satisfying $PP^T v = b$. After some algebraic operations, we obtain that

$$v = \frac{\gamma_1 - M|c_N|}{\sum_{i=1}^{N-1} |c_i|-|c_N|}$$

and

$$x_+ = \begin{bmatrix} |c_1|-|c_N| \\ |c_2|-|c_N| \\ \vdots \\ |c_{N-1}|-|c_N| \end{bmatrix} \frac{\gamma_1 - M|c_N|}{\sum_{i=1}^{N-1} (|c_i|-|c_N|)^2}.$$

Thus the scaling factors corresponding to the optimal solution are

$$a_j = \frac{|c_j|-|c_N|}{\sum_{i=1}^{N-1}(|c_i|-|c_N|)^2}(\gamma_1 - M|c_N|), \quad j=1,2,\ldots,N-1,$$

$$a_N = M - \sum_{i=1}^{N-1} a_j = \frac{\sum_{i=1}^{N-1}(|c_i|-|c_N|)(\gamma_1 - M|c_i|)}{\sum_{i=1}^{N-1}(|c_i|-|c_N|)^2}.$$

(34)

Suppose all scaling factors from (34) are positive, we say the optimal solution does exist and (32) is satisfied by this optimal solution. If it is not, i.e., some of them are negative or zero, we say the optimal solution does not exist and we seeking for a suboptimal solution. Assume that only the factor $a_k$ becomes non-positive, this occurs when the signs of $|c_k|-|c_N|$ and $\gamma_1 - M|c_N|$ in (34) are opposite. Since

$$\gamma_1 - M|c_N| = \gamma_1 - M|c_k| + M(|c_k|-|c_N|)$$

then either $\gamma_1/M < |c_N| < |c_k|$ or $|c_k| < |c_N| < \gamma_1/M$ holds. At the same time, all other factors are positive, i.e., $|c_j| - |c_N|$ has the same sign as $\gamma_1 - M|c_N|$ in (34) for $1 \leq j \neq k \leq N$. Thus either $|c_k|$ is the largest or smallest absolute value in the DWT coefficients depending on $\gamma_1/M < |c_N|$ or $|c_N| < \gamma_1/M$. In order not to discredit the SNR, we need to keep this DWT coefficient term to be un-scaled, i.e., $a_k = 1$ is selected. And we resolve (32) for the optimal solution with the other $N-1$ scaling factors. If more factors are not positive, we seek for the other factors to maximize the SNR. Therefore, this optimization process will be performed iteratively.

For the case to embed the bit "$0 \in B$", we can replace the parameter $\gamma_1$ by $\gamma_0$ and obtain the corresponding values for $a_1, a_2, \ldots, a_N$ by the similar analysis as just discussed.

C. *The scaling factor's invariance to amplitude scaling*

Let $\bar{S}(t)$ and $\tilde{S}(t)$ be the signals prior to and after amplitude scaling. Since these two signals form scale multiple of each other, we have

$$\tilde{S}(t) = \tau \cdot \bar{S}(t) \tag{35}$$

where $\tau$ being the scaling factor and the relation in (3) implies

$$\begin{aligned}\tilde{S}(t) = \tau \bar{S}(t) &= \tau \sum_k \bar{c}_{j_0}(k)\varphi_{j_0,k}(t) + \tau \sum_k \sum_{j=j_0}^{\infty} \bar{d}_j(k)\psi_{j,k}(t) \\ &= \sum_k \tau \bar{c}_{j_0}(k)\varphi_{j_0,k}(t) + \sum_k \sum_{j=j_0}^{\infty} \tau \bar{d}_j(k)\psi_{j,k}(t) \\ &= \sum_k \tilde{c}_{j_0}(k)\varphi_{j_0,k}(t) + \sum_k \sum_{j=j_0}^{\infty} \tilde{d}_j(k)\psi_{j,k}(t). \end{aligned} \tag{36}$$

Throughout this study, the host digital audio signal $S(n)$, $n \in \mathbb{N}$, which denotes the samples of the original audio signal $S(t)$ at the $n$-th sample time, is cut into segments on which DWT are preformed. Let $\bar{C}_N = \{|\bar{c}_i| : 0 \leq i \leq N-1\}$ and $\tilde{C}_N = \{|\tilde{c}_i| : 0 \leq i \leq N-1\}$ be the sets of low-frequency DWT coefficients of $\bar{S}(n)$ and $\tilde{S}(n)$, which correspond to the digital audio signals prior to and after amplitude scaling attack, respectively. The relation in (37) implies that

$$\tilde{a}_j = \frac{|\tilde{c}_j| - |\tilde{c}_N|}{\sum_{i=1}^{N-1}(|\tilde{c}_i| - |\tilde{c}_N|)^2}(\gamma_1 - M|\tilde{c}_N|) = \frac{|\tau\bar{c}_j| - |\tau\bar{c}_N|}{\sum_{i=1}^{N-1}(|\tau\bar{c}_i| - |\tau\bar{c}_N|)^2}(\gamma_1 - M|\tau\bar{c}_N|) = \frac{|\bar{c}_j| - |\bar{c}_N|}{\sum_{i=1}^{N-1}(|\bar{c}_i| - |\bar{c}_N|)^2}(\gamma_1 - M|\bar{c}_N|) = \bar{a}_j,$$

where $j = 1, 2, \ldots, N-1$, and

$$\tilde{a}_N = M - \sum_{i=1}^{N-1}\tilde{a}_j = \frac{\sum_{i=1}^{N-1}(|\tilde{c}_i| - |\tilde{c}_N|)(\gamma_1 - M|\tilde{c}_i|)}{\sum_{i=1}^{N-1}(|\tilde{c}_i| - |\tilde{c}_N|)^2} = \frac{\sum_{i=1}^{N-1}(|\tau\bar{c}_i| - |\tau\bar{c}_N|)(\gamma_1 - M|\tau\bar{c}_i|)}{\sum_{i=1}^{N-1}(\tau|\bar{c}_i| - |\tau\bar{c}_N|)^2} = \bar{a}_N.$$

It is worth to mention that the invariance of scaling factors indicates the nice feature which is robust to amplitude scaling. Hence we use scaling factors to be the second watermarks in testing amplitude scaling.

## IV. EXPERIMENTAL RESULTS

This section shows the performance of the proposed optimization-based amplitude quantization for audio watermarking. We first split the original audio into four segments. By applying seven-level DWT to each segment, we embed the synchronization code and watermarks into the lowest-frequency sub-band. To contrast the SNR of single-coefficient quantization sizes $Q = 6500$ in reference paper [2], we select $M = N$ for our experiment and the amplitude quantization sizes $Q$ are set to 26000 and 52000 for $N=4$ and $N=8$. There are four kinds of audio signals which are love song, symphony, dance, and folklore respectively for our testing. These audio signals are 16-bit mono-type with sampling rate 44.1 kHz.

Table I shows the domain, SNR, and the embedding capacity. The embedding capacity is measured with 11.6 seconds audio signal. One can see that the proposed method keeps high SNR due to the fact that every coefficient is optimized. Moreover, the relationship between the SNR and the quantization size $Q$ for case $N=4$ is shown in Fig. 7. It indicates that the SNR is decreasing when $Q$ is increasing.

After the embedding process of watermarking, we apply some attacks in order to test the robustness which is measured by

$$\text{BER} = \frac{B_{error}}{B_{total}} \times 100\%,$$

where $B_{error}$ and $B_{total}$ denote the number of error binary bits and the number of total binary bits. These attacks include 4 types: (1) re-sampling, (2) low-pass filtering, (3) time scaling, and (4) amplitude scaling. They are introduced as follows.

(1) *Re-sampling:* The watermarked audio was down-sampled from 44.1kHz to 22.05kHz, and then back to 44.1kHz by interpolation. Similarly, the sampling rate was varied from 44.1kHz to 11.025kHz, and 8kHz, and then back to 44.1kHz. Tables II-V show the experiments of these re-sampling processes. Based on these results, the proposed scheme is more robust than the one proposed by Wu *et al.* [2]. In addition, the proposed method was found to have similar robustness with the methods proposed by Xiang *et al.* [16].

(2) *Low-pass filtering:* Tables VI-IX show the effect of adopting a low-pass filter with the cutoff frequency 3 kHz. The proposed method has better robustness than the method proposed by Wu *et al.* [2]. However, the proposed method has slightly lower robustness than the method proposed by Xiang *et al.* [16].

(3) *Amplitude scaling:* Since a big scaling factor will result in saturation, we set the scaling factor $\alpha$ as 0.5, 0.8, 1.1, and 1.2. Based on the invariance of the optimal scaling factors, Tables X-XIII show the good experimental results with this attack.

(4) *Time scaling:* The watermarked audios are scaled by −5%, -2%, 2%, and 5%. Tables XI-XIV show the experiments of this attack. They indicate that the method of [2] has lower robustness than ours. Moreover, Figure 8 shows the relationship between the BER and quantization size Q for the four songs after time scaling -5% ($N=4$). It shows that the BER has similar results when the quantization size Q increases.

## V. CONCLUSION

This study presents an optimization-based amplitude quantization scheme with optimal scale on wavelet coefficients. To enhance the robustness, the watermarks are embedded into DWT lowest-frequency coefficients. These coefficients and the scale factors of these coefficients are both optimized. Experimental results show that the embedded audio has high

SNR and the hidden information has better robustness against signal processing and attacks, such as re-sampling, low-pass filtering, and amplitude scaling. In the future work, we will aim to propose an invariant feature in the wavelet domain to resist time scaling.

## REFERENCES

[1] S. Katzenbeisser and F. A. P. Petitcolas, Eds., *Information Hiding Techniques for Steganography and Digital Watermarking*: Artech House, Inc., 2000.

[2] S. Wu, J. Huang, D. Huang, and Y. Q. Shi, "Efficiently self-synchronized audio watermarking for assure audio data transmission," *IEEE. Transactions on Broadcasting*, vol.51, no.1, pp. 69-76, March 2005.

[3] M. A. Gerzon and P. G. Graven, "A high-rate buried-data channel for audio CD," *Journal of the Audio Engineering Society*, vol. 43, no. 1/2, pp. 3-22, 1995.

[4] D. Gruhl "Echo hiding," in *Proceedings of the 1st Information Hiding Workshop*, *LNCS*, vol. 1174, Berlin, Germany, pp. 295-315, 1996.

[5] H. J. Kim, "Audio watermarking techniques," *Pacific Rim Workshop on Digital Steganography*, Kyushu Institute of Technology, Kitakyushu, Japan, July 3-4, 2003

[6] H. Alaryani and A. Youssef, "A novel audio watermarking technique based on frequency components," *Proceedings of the Seventh IEEE International Symposium on Multimedia*, 2005.

[7] W. N. Lie and L. C. Chang, "Robust and high-quality time-domain audio watermarking based on low-frequency amplitude modification," *IEEE. Transactions on Multimedia*, vol.8, no.1, pp. 46-59, February 2006.

[8] L. Boney, A. H. Tewfik, and K. N. Handy, "Digital watermarks for audio signals," in *Proceeding 3rd IEEE International Conference on Multimedia Computing and Systems*, pp. 473-480, 1996.

[9] T. Ciloglu and S. U. Karaaslan, " An improved all-pass watermarking scheme for speech and audio," in *Proceeding 3 rd IEEE International Conference on Multimedia Computing and Exposition (ICME)*, pp. 1017-1020, 2000.

[10] H. Kim, "Stochastic model based audio watermark whitening filter for improve detection," in *Proceeding 3 rd IEEE International Conference on Acoustics, Speech and Signal Processing (ICASSP)*, pp. 1971-1974, 2000.

[11] P. Bassia, I. Pitas, and N. Nikolaidis, "Robust audio watermarking in the time domain," *IEEE Transaction on Multimedia*, vol. 3, no 2, pp.232-241, Jun. 2001.

[12] B. S. Ko, R. Nishimura, and Y. Suzuki, "Time-spread echo method for digital audio watermarking using PN sequence," in *Proceeding IEEE International Conference on Acoustics, Speech and Signal Processing*, vol. II, pp. 2001-2004, 2002.

[13] J. Huang, Y. Wang, and Y. Q. Shi, "A blind audio watermarking algorithm with

self-synchronization," in *Proceeding IEEE International Symposium on Circuits and Systems*, vol. 3, pp. 627-630, 2002.

[14] W. Bender, D. Gruhl, and N. Morimoto, "Techniques for data hiding," *IBM Syst. J.*, vol.35, no.3/4, pp. 131-336, 1996.

[15] C. P. Wu, P. C. Su, and C. C. J. Kuo, "Robust frequency domain audio watermarking based on audio content analysis," in *Proceeding International Symposium on Multimedia Information Processing*, *ISMIP*, PP. 37-45, 1999.

[16] S. Xiang, and J. Huang, "Robust audio watermarking against the D/A and A/D conversions," arXiv:0707.0397v1, http://arxiv.org/abs/0707.0397 (2007).

[17] S.-T. Chen, H.-N. Huang, C.-J. Chen, and G.-D. Wu, "Energy-proportion based scheme for audio watermarking," *IET Proceedings on Signal Processing*, vol. 4, no. 5, pp. 576-587, 2010.

[18] S.-T. Chen, G.-D. Wu, and H.-N. Huang, "Wavelet-domain audio watermarking scheme using optimization-based quantization," *IET Proceedings on Signal Processing*, vol. 4, no. 6, pp. 720-727, 2010.

[19] S. Mallat, "A theory for multiresolution signal decomposition: the wavelet representation," *IEEE. Transaction on Pattern Anal. And Machine Intel.*, vol.11, pp.674-693, July 1989.

[20] F. L. Lewis, *Optimal Control.* New York: John Wiley and Sons, 1986.

[21] R. G. Bartle, *The Elements of Real Analysis.* New York: Wiley, second ed., 1976.

[22] C. S. Burrus, R. A. Gopinath, and H. Gao, *Introduction to Wavelet Theory and Its Application.* New Jersey: Prentice-Hall 1998.

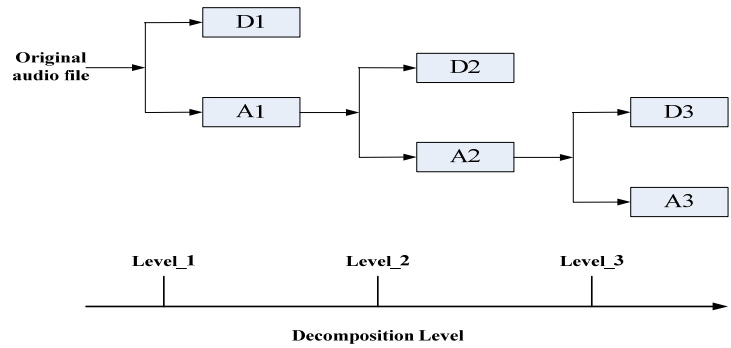

Fig. 1. DWT

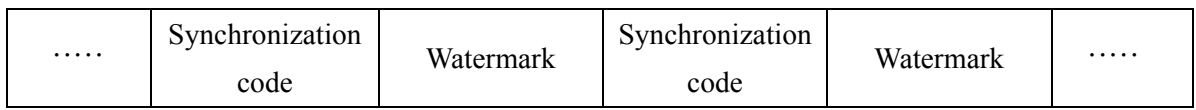

Fig. 2. Structure of synchronization codes and watermarks

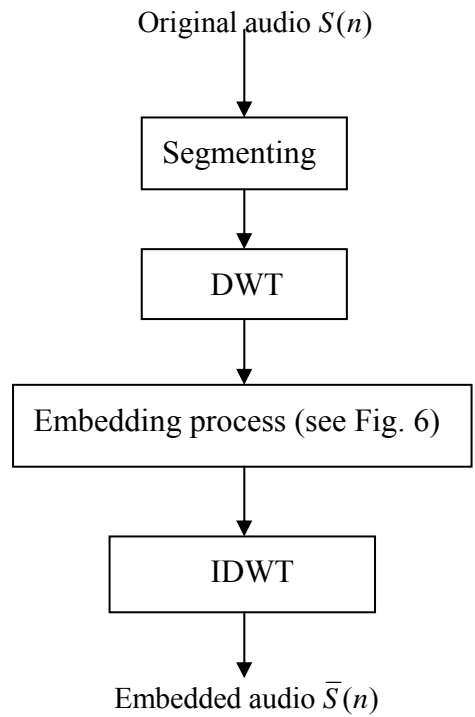

Fig. 3. Watermark embedding.

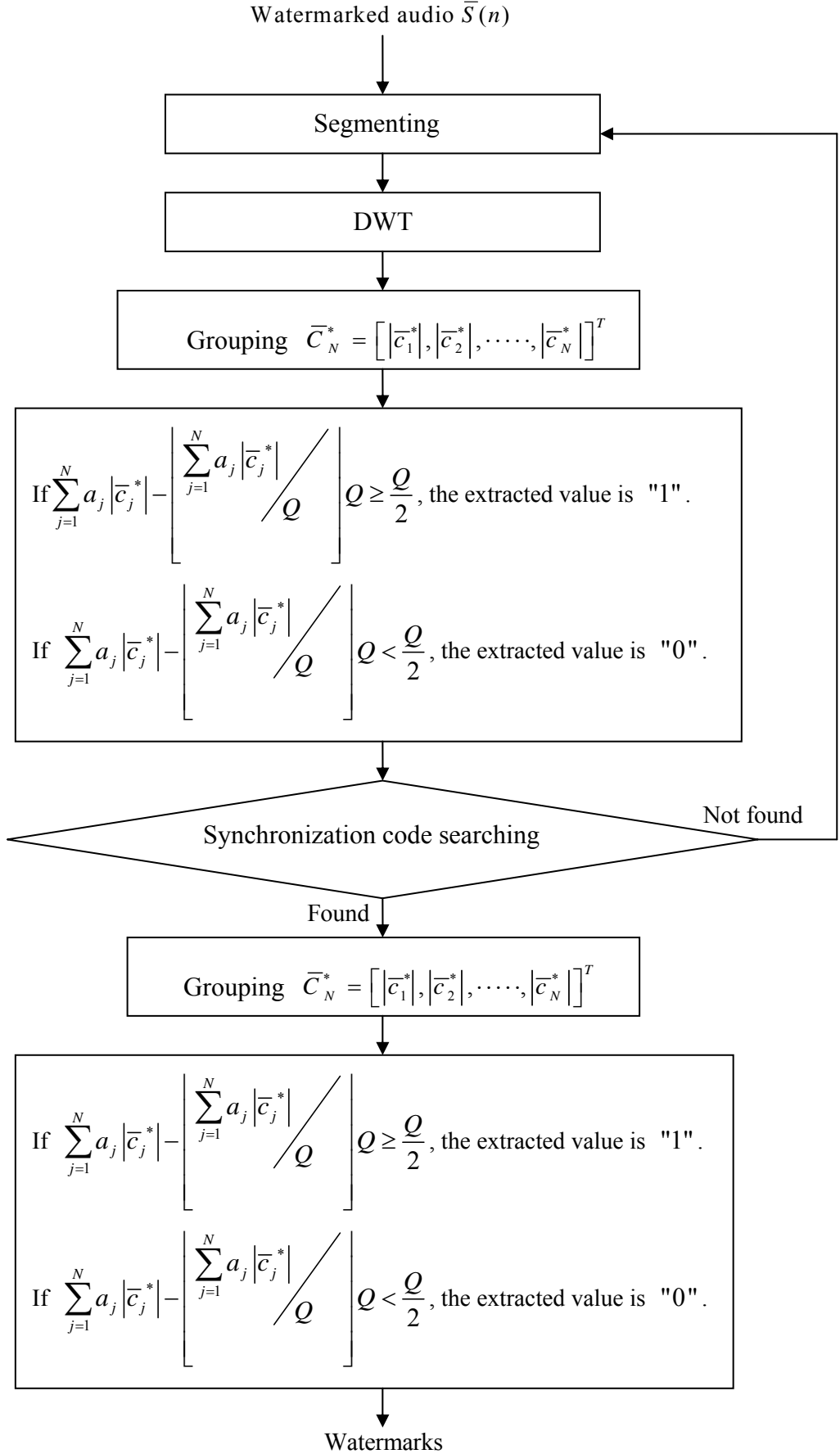

Fig. 4. Watermark extraction.

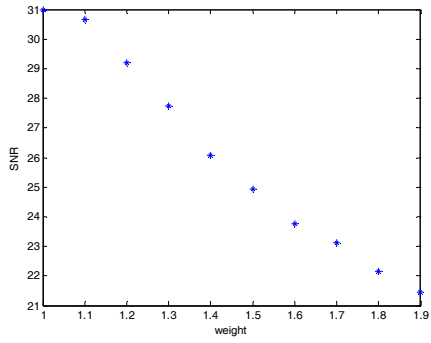
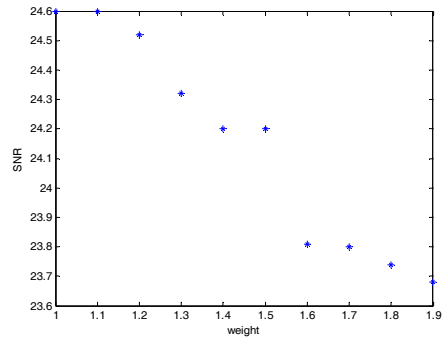

(a) love song　　　　　　　　　　　　　(b) symphony

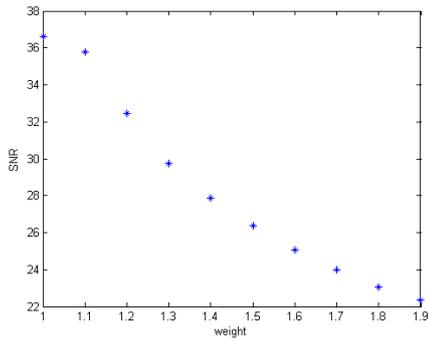
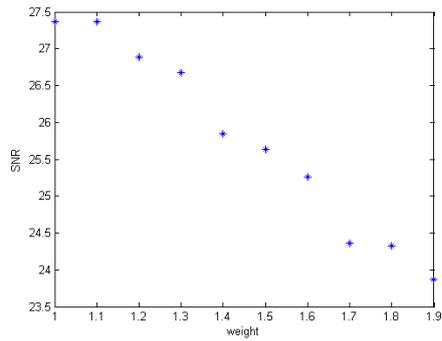

(c) dance　　　　　　　　　　　　　　(d) folklore

Fig. 5. Variation of the SNR to $a_2$ for four types of song when *N*=4 and *M*=4.

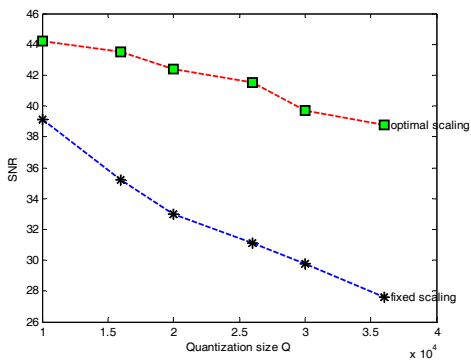
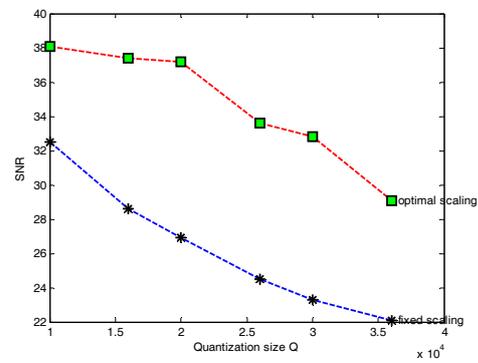

(a) love song　　　　　　　　　　　　　(b) symphony

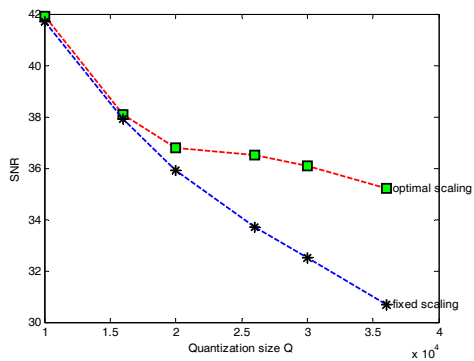
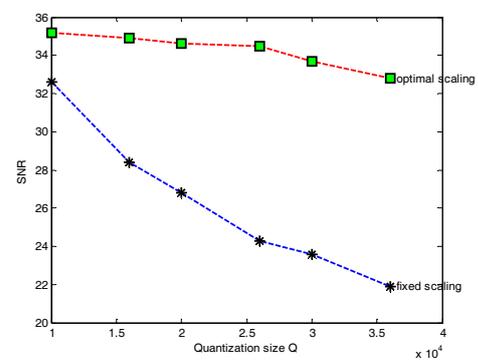

(c) dance　　　　　　　　　　　　　　(d) folklore

Fig. 7. The relationship between the SNR and quantization size Q for the optimal scaling method and fixed scaling method under *N*=4.

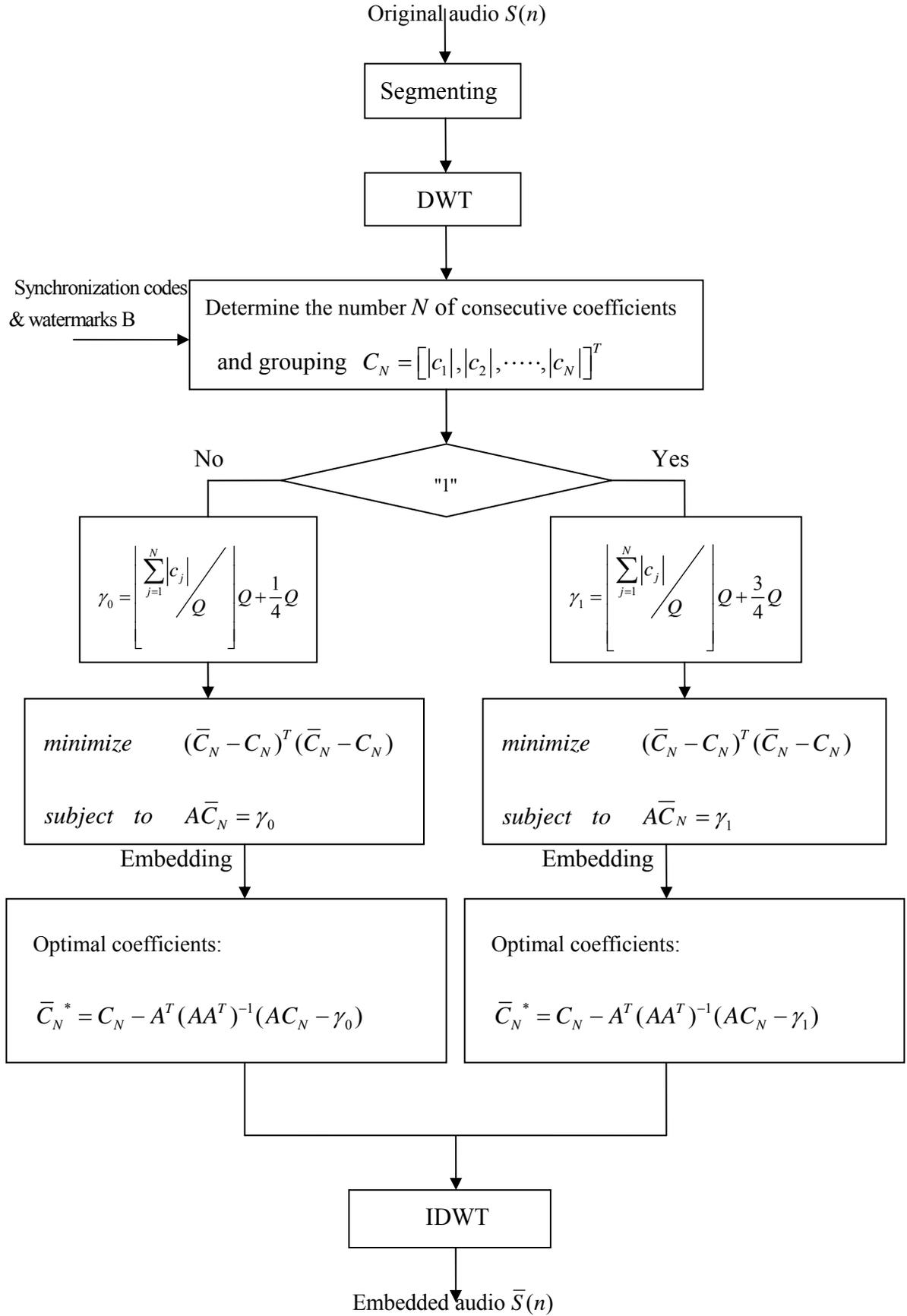

Fig. 6. Embedding process

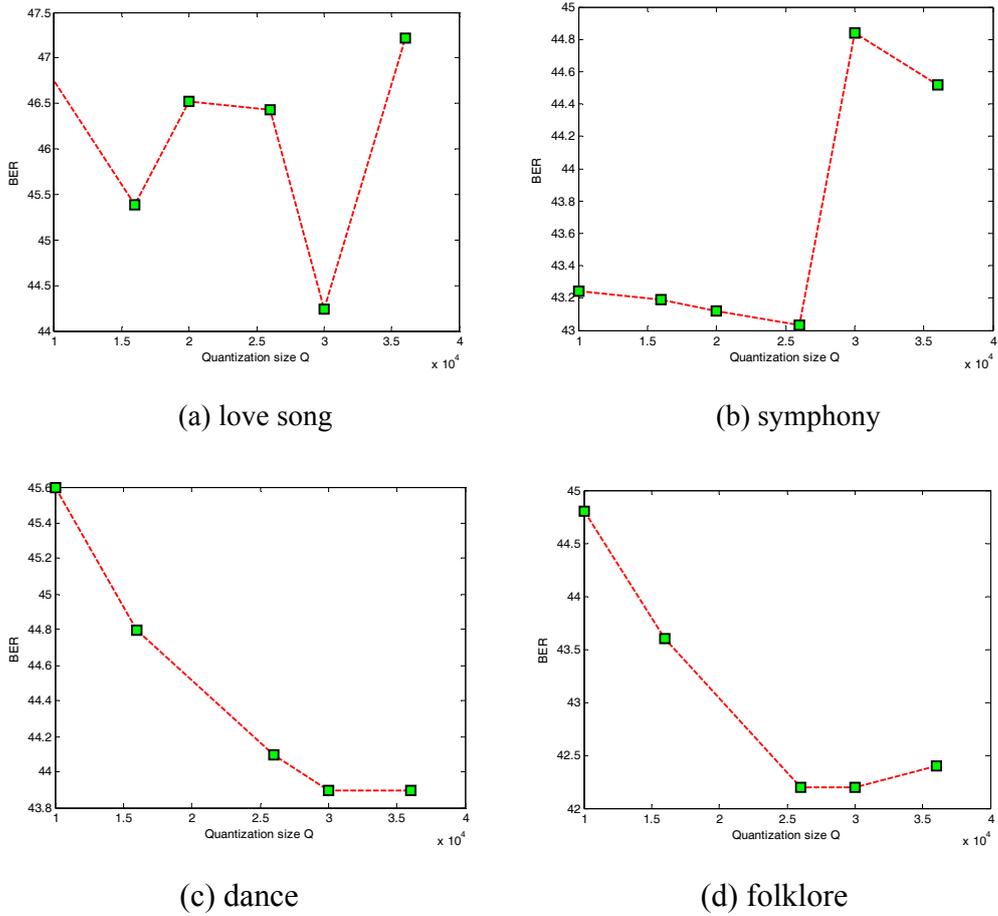

(a) love song

(b) symphony

(c) dance

(d) folklore

Fig. 8. The relationship between the BER and quantization size Q for two song "love song" and "symphony" after time scaling -5% ($N=4$).

**TABLE I.** DOMAIN, SNR, AND EMBEDDING CAPACITY.

|  |  | Domain | SNR(dB) |  | Embedding capacity |
|---|---|---|---|---|---|
| Reference paper [2] |  | DWT 8-level | 33.6(love song) 36.4(dance) | 26.9(symphony) 27.2(folklore) | 2000bits/ 11.6seconds |
| Reference paper [16] |  | DWT 7-level | 17.8(love song) 16.5(dance) | 18.2(symphony) 17.4(folklore) | 666bits/ 11.6seconds |
| Proposed Method | $N=4$ | DWT 7-level | 41.5(love song) 36.5(dance) | 33.6(symphony) 34.5(folklore) | 1000bits/ 11.6seconds |
|  | $N=8$ | DWT 7-level | 41.4(love song) 36.4(dance) | 33.9(symphony) 34.4(folklore) | 500bits/ 11.6seconds |

TABLE II. RE-SAMPLING (love song).

| Re-sampling Rate(Hz) | | 22050 | 11025 | 8000 |
|---|---|---|---|---|
| Reference paper[2] BER(%) | | 7.75 | 19.04 | 18.28 |
| Reference paper[16] BER(%) | | 1.24 | 7.82 | 8.25 |
| Proposed method BER(%) | $N = 4$ | 1.30 | 8.43 | 8.46 |
| | $N = 8$ | 0.20 | 5.41 | 5.02 |

TABLE III. RE-SAMPLING (symphony).

| Re-sampling Rate(Hz) | | 22050 | 11025 | 8000 |
|---|---|---|---|---|
| Reference paper[2] BER(%) | | 0.80 | 1.28 | 1.32 |
| Reference paper[16] BER(%) | | 0.42 | 1.02 | 1.26 |
| Proposed method BER(%) | $N = 4$ | 0.51 | 1.63 | 1.62 |
| | $N = 8$ | 0.40 | 1.00 | 1.00 |

TABLE IV. RE-SAMPLING (dance).

| Re-sampling Rate(Hz) | | 22050 | 11025 | 8000 |
|---|---|---|---|---|
| Reference paper[2] BER(%) | | 14.25 | 28.35 | 27.55 |
| Reference paper[16] BER(%) | | 3.21 | 8.92 | 9.03 |
| Proposed method BER(%) | $N = 4$ | 2.36 | 8.01 | 8.01 |
| | $N = 8$ | 1.21 | 3.23 | 2.66 |

TABLE V. RE-SAMPLING (folklore).

| Re-sampling Rate(Hz) | | 22050 | 11025 | 8000 |
|---|---|---|---|---|
| Reference paper[2] BER(%) | | 0.85 | 10.65 | 10.20 |
| Reference paper[16] BER(%) | | 0.44 | 2.06 | 2.42 |
| Proposed method BER(%) | $N = 4$ | 0.20 | 1.24 | 1.26 |
| | $N = 8$ | 0 | 0 | 0.01 |

TABLE VI. LOW-PASS FILTERING.

| Audio type | | love song | symphony | dance | folklore |
|---|---|---|---|---|---|
| Reference paper[2]BER(%) | | 40.64 | 23.24 | 36.24 | 30.30 |
| Reference paper[16] BER(%) | | 20.62 | 7.81 | 21.08 | 10.54 |
| Proposed method BER(%) | $N = 4$ | 33.40 | 6.38 | 33.02 | 20.23 |
| | $N = 8$ | 24.60 | 2.24 | 27.56 | 11.80 |

**TABLE VII.** AMPLITUDE SCALING (love song).

| Scaling factor | | 0.5 | 0.8 | 1.1 | 1.2 |
|---|---|---|---|---|---|
| Reference paper[2] BER(%) | | 51.62 | 43.02 | 31.32 | 40.54 |
| Reference paper[16] BER(%) | | 2.47 | 2.03 | 1.98 | 1.99 |
| Proposed method BER(%) | $N = 4$ | 1.25 | 0.85 | 0.83 | 0.91 |
| | $N = 8$ | 1.22 | 0.82 | 0.82 | 0.89 |

**TABLE VIII.** AMPLITUDE SCALING (symphony).

| Scaling factor | | 0.5 | 0.8 | 1.1 | 1.2 |
|---|---|---|---|---|---|
| Reference paper[2] BER(%) | | 51.48 | 17.93 | 8.63 | 13.76 |
| Reference paper[16] BER(%) | | 2.06 | 2.01 | 1.84 | 1.88 |
| Proposed method BER(%) | $N = 4$ | 1.12 | 0.86 | 0.90 | 0.96 |
| | $N = 8$ | 1.10 | 0.85 | 0.89 | 0.96 |

**TABLE IX.** AMPLITUDE SCALING (dance).

| Scaling factor | | 0.5 | 0.8 | 1.1 | 1.2 |
|---|---|---|---|---|---|
| Reference paper[2] BER(%) | | 47.26 | 41.61 | 41.06 | 42.86 |
| Reference paper[16] BER(%) | | 4.65 | 4.46 | 4.12 | 4.32 |
| Proposed method BER(%) | $N = 4$ | 2.01 | 1.58 | 1.02 | 1.94 |
| | $N = 8$ | 1.98 | 1.57 | 1.02 | 1.94 |

**TABLE X.** AMPLITUDE SCALING (folklore).

| Scaling factor | | 0.5 | 0.8 | 1.1 | 1.2 |
|---|---|---|---|---|---|
| Reference paper[2] BER(%) | | 51.35 | 36.97 | 22.52 | 31.21 |
| Reference paper[16] BER(%) | | 1.68 | 1.44 | 1.22 | 1.56 |
| Proposed method BER(%) | $N = 4$ | 1.03 | 0.82 | 0.82 | 0.86 |
| | $N = 8$ | 1.02 | 0.81 | 0.81 | 0.85 |

**TABLE XI.** TIME SCALING (love song).

| Time scaling (%) | | -5 | -2 | 2 | 5 |
|---|---|---|---|---|---|
| Reference paper[2] BER(%) | | 50.14 | 48.92 | 50.23 | 50.84 |
| Reference paper[16] BER(%) | | 47.24 | 43.21 | 41.06 | 44.56 |
| Proposed method BER(%) | $N = 4$ | 46.45 | 40.05 | 44.83 | 46.58 |
| | $N = 8$ | 46.45 | 36.78 | 44.42 | 44.35 |

**TABLE XII.** TIME SCALING (symphony).

| Time scaling (%) | | -5 | -2 | 2 | 5 |
|---|---|---|---|---|---|
| Reference paper[2] BER(%) | | 51.81 | 49.23 | 51.13 | 52.36 |
| Reference paper[16] BER(%) | | 42.63 | 42.14 | 45.87 | 45.43 |
| Proposed method BER(%) | $N = 4$ | 43.03 | 37.82 | 46.29 | 46.26 |
| | $N = 8$ | 41.42 | 37.40 | 46.29 | 46.14 |

**TABLE XIII.** TIME SCALING (dance).

| Time scaling (%) | | -5 | -2 | 2 | 5 |
|---|---|---|---|---|---|
| Reference paper[2] BER(%) | | 51.62 | 49.61 | 51.75 | 52.86 |
| Reference paper[16] BER(%) | | 43.76 | 43.22 | 43.58 | 43.23 |
| Proposed method BER(%) | $N = 4$ | 44.11 | 39.58 | 47.92 | 47.98 |
| | $N = 8$ | 42.25 | 39.14 | 47.36 | 47.24 |

**TABLE XIV.** TIME SCALING (folklore).

| Time scaling (%) | | -5 | -2 | 2 | 5 |
|---|---|---|---|---|---|
| Reference paper[2] BER(%) | | 51.65 | 49.67 | 51.72 | 52.81 |
| Reference paper[16] BER(%) | | 43.22 | 41.74 | 41.23 | 42.16 |
| Proposed method BER(%) | $N = 4$ | 42.23 | 38.22 | 46.43 | 47.36 |
| | $N = 8$ | 41.25 | 38.21 | 46.38 | 46.61 |